\documentclass[prl,aps,epsf,amsfonts,floats,twocolumn,amssymb,amsmath,groupedaddress,showpacs,floatfix,nofootinbib]{revtex4}
\usepackage{graphicx}
\usepackage{epsf}

\usepackage[]{latexsym}
\usepackage{bm}

\newcommand{\be}{\begin{equation}}\newcommand{\ee}{\end{equation}}
\newcommand{\bea}{\begin{eqnarray}}\newcommand{\eea}{\end{eqnarray}}
\newcommand{\brr}{\begin{array}}\newcommand{\err}{\end{array}}
\newcommand{\bit}{\begin{itemize}}\newcommand{\eit}{\end{itemize}}
\newcommand{\ben}{\begin{enumerate}}\newcommand{\een}{\end{enumerate}}

\newcommand{\ba}{\begin{array}}
\newcommand{\ea}{\end{array}}

\def\non{\nonumber}
\def\rar{\rightarrow}

\def\1{{_{1}}}\def\2{{_{2}}}

\def\noHe0{:\;\!\!\;\!\!:H_e(0):\;\!\!\;\!\!:}
\def\noHm0{:\;\!\!\;\!\!:H_\mu(0):\;\!\!\;\!\!:}

\def\non{\nonumber}

\def\rar{\rightarrow}

\def\1{{_{1}}}\def\2{{_{2}}}

\begin{document}

\title{Spontaneous supersymmetry breaking probed by geometric invariants}

\author{ Antonio Capolupo${}^{\natural}$\footnote{Corresponding author: tel.: +39 089 968213; e-mail address: capolupo@sa.infn.it (A.Capolupo).}}
\author{ Giuseppe Vitiello${}^{\flat}$}
 \affiliation{${}^{\natural}$ Dipartimento di Ingegneria Industriale,
  Universit\'a di Salerno, Fisciano (SA) - 84084, Italy}
   \affiliation{${}^{\flat}$
  Dipartimento di Fisica E.R.Caianiello
  Universit\'a di Salerno, and INFN Gruppo collegato di Salerno, Fisciano (SA) - 84084, Italy}


\begin{abstract}

The presence of the Aharonov-Anandan invariant in phenomena in which vacuum condensates are physically relevant can help to reveal the spontaneous supersymmetry  breaking induced by condensates.
The analysis is presented in the case of the Wess--Zumino model. The manifestation of the Aharonov-Anandan invariant of atoms and their superpartners, generated at non-zero temperature, could reveal the signature of SUSY violation in
 a recently proposed experimental setup based on an optical lattice in which SUSY is broken at
non-zero temperature.

\end{abstract}

\maketitle

The idea of supersymmetry (SUSY) had a huge impact on physics in the past years and much of the research has been geared at the detection of the superpartners,   which according to SUSY are associated with the ordinary particles we know. However, no evidence of the existence of superpartners has been obtained till now.  The fact that the superpartners are not degenerate with ordinary particles at the scales we can currently probe means that, if SUSY exists as a fundamental symmetry, it must be spontaneously broken. As a consequence,  a large part of the research activity has been devoted to the study of SUSY breaking.\\
\indent It has been recently shown \cite{Capolupo:2012vf} that, in several phenomena  in a supersymmetric context, with particular reference to the free Wess-Zumino model \cite{Wess:1973kz}, SUSY is broken by the vacuum condensate shifting the vacuum energy density to a non-zero value.  Examples include QFT in  external fields, like Unruh \cite{Unruh:1976db} and  Schwinger effects \cite{Schwinger:1951nm}, condensed matter physics, such as discussed in connection with BCS theory of superconductivity \cite{Bardeen:1957mv}, graphene physics \cite{Iorio:2010pv} and Thermo Field Dynamics \cite{Takahashi:1974zn,Umezawa:1993yq,Vit}, particle physics and cosmology, flavor mixing (see, e.g. \cite{Blasone:2002jv,Blasone:2001du} and refs. therein cited), quantization of dissipative systems \cite{Celeghini:1991yv} and dark energy \cite{Capolupo:2006et,Capolupo2} even though the dark energy problem is far from being settled.  In all these cases, vacuum condensates can be effectively described by using Bogoliubov transformations.
\\
\indent On the other hand, the presence of Aharonov-Anandan invariant (AAI) \cite{Anandan:1990fq} has been revealed to be an interesting feature in the study of phenomena characterized by a vacuum condensate  \cite{Capolupo:2013fasi}.
\\
\indent In this paper we show that in a supersymmetric context, the AAI could represent a valid instrument to reveal the SUSY violation induced by condensates.

In the following we summarize the main results of \cite{Capolupo:2012vf} and then we show that SUSY breaking could be revealed by means of AAIs.
\\
\indent {\it Vacuum condensate  and SUSY breaking} -
We consider, for simplicity,  the free Wess--Zumino model Lagrangian \cite{Wess:1973kz}  which is invariant under SUSY. This leads us to the qualitative understanding of the SUSY breaking in more complicate systems.\\
\indent In order not to break SUSY explicitly, we study the vacuum condensation induced by Bogoliubov transformations acting simultaneously and with the same parameters on the bosonic and  the fermionic degrees of freedom.  The condensate shifts the vacuum energy density to a nonvanishing value and, as well known, this is a sufficient condition for the spontaneous breaking of SUSY to occur  \cite{Witten:1981nf}.

The Lagrangian of the free Wess--Zumino model is given by
\bea \label{WS}\non
\mathcal{L} &=& \frac{i}{2} \,\bar{\psi}\gamma_{\mu}\partial^{\mu}\psi + \frac{1}{2}\,\partial_{\mu}S\partial^{\mu}S + \frac{1}{2}\,\partial_{\mu}P\partial^{\mu}P
\\&-&  \frac{m}{2} \, \bar{\psi}\psi - \frac{m^2}{2}\, (S^2 + P^2),
\eea
where $\psi$ is a Majorana spinor field, $S$ is a scalar field and $P$ is a pseudoscalar field. This Lagrangian is invariant under supersymmetry transformations \cite{Wess:1973kz}.
Let  $\alpha^r_{\mathbf{k}}$, $b_{\mathbf{k}}$ and $c_{\mathbf{k}}$ denote the annihilation operators of the quantized fields
$\psi$, $S$ and $P$, respectively, which annihilate the vacuum $|0\rangle \equiv |0\rangle_{\psi}\otimes|0\rangle_S \otimes |0\rangle_P$.
Consider then the Bogoliubov transformations of the ladder operators corresponding to all three fields. In full generality we consider time-dependent Bogoliubov transformations (their specific nature is dictated by the physical phenomena and system under study):
\bea\label{Bog1}
\tilde{\alpha}^r_{\mathbf{k}}(\xi, t) &=& U^{\psi}_{\mathbf{k}}(\xi,t) \, \alpha^r_{\mathbf{k}}(t) + V^{\psi}_{-\mathbf{k}}(\xi,t)  \, \alpha^{r\dagger}_{-\mathbf{k}}( t)\,,
\\[2mm]\label{Bog2}
\tilde{b}_{\mathbf{k}}(\eta,  t) &=& U^{S}_{\mathbf{k}}(\eta,t)  \, b_{\mathbf{k}}(t) - V^{S}_{-\mathbf{k}}(\eta,t)  \,b^{\dagger}_{-\mathbf{k}}(t)\,,
\\
[2mm]
\tilde{c}_{\mathbf{k}}(\eta,  t) &=& U^{P}_{\mathbf{k}}(\eta,t) \, c_{\mathbf{k}}(t) - V^{P}_{-\mathbf{k}} (\eta,t) \,c^{\dagger}_{-\mathbf{k}}(t)\,,
\label{Bog3}
\eea
where the coefficients of scalar and pseudoscalar bosons are the same, $U^{S}_{\mathbf{k}} =U^{P}_{\mathbf{k}} $ and $V^{S}_{\mathbf{k}} =V^{P}_{\mathbf{k}} $. Denote them by $U^{B}_{\mathbf{k}} $ and $V^{B}_{\mathbf{k}} $, respectively. We do not specify further the explicit form of these  coefficients $U_{\mathbf{k}}$ and $V_{\mathbf{k}}$ except for the fact that $U_{\mathbf{k}} = U_{- \mathbf{k}}$ and $V_{\mathbf{k}} = \pm V_{- \mathbf{k}}$, with $\pm$ for bosons and fermions, respectively, and they satisfy the relation
$|U^{B}_{\mathbf{k}} |^2 - |V^{B}_{\mathbf{k}} |^2=1$  for  (boson) $S$ and $P$, and $|U^{\psi}_{\mathbf{k}} |^2 + |V^{\psi}_{\mathbf{k}} |^2=1$  for (fermion) $\psi$, which guarantees that the transformations are  canonical.
For notational simplicity we write $U^{\psi}_{\mathbf{k}}\equiv U^{\psi}_{\mathbf{k}}(\xi,t) $, $U^{B}_{\mathbf{k}}\equiv U^{B}_{\mathbf{k}}(\eta,t) $, etc.

At any given time $t$ the transformations (\ref{Bog1})--(\ref{Bog3}) can be written as:
$
\tilde{\alpha}^r_{\mathbf{k}}( \xi,  t) = J^{-1} (\xi,  t)\,\alpha^r_{\mathbf{k}}(t) J(\xi,  t)\,,
$
and similar relations for the other  operators. We do not report here the explicit form of the generators $J$s since it does not affect our discussion  (it is given, e.g., in \cite{Capolupo:2012vf}).
The operators $\tilde{\alpha}^r_{\mathbf{k}}( \xi,  t)$, $\tilde{b}_{\mathbf{k}}(\eta,  t)$ and $\tilde{c}_{\mathbf{k}}(\eta,  t)$ annihilate  the (tensor product) vacuum
\bea
|\tilde{0}(\xi, \eta,   t)\rangle \equiv |\tilde{0}(\xi,   t)\rangle_{\psi}\otimes |\tilde{0}(\eta,  t)\rangle_{S}\otimes |\tilde{0}(\eta,   t)\rangle_{P},
\eea
 where
$
|\tilde{0}(\xi,  t)\rangle_{\psi} = J^{-1}_{\psi}(\xi,  t)|0\rangle_{\psi} $,
$ |\tilde{0}(\eta,  t)\rangle_{S} = J^{-1}_{S}(\eta,  t)|0\rangle_{S}$ and
$  |\tilde{0}(\eta,   t)\rangle_{P} = J^{-1}_{P}(\eta,  t)|0\rangle_{P}
$, with obvious notation for the $J_{\psi}$, $J_{S}$ and $J_{P}$ generators.
We write in a compact notation
$
|\tilde{0}( t)\rangle \equiv  |\tilde{0}(\xi, \eta,  t)\rangle = J^{-1}(\xi,\eta,  t)|0\rangle\,.
$

In a similar way, the Bogoliubov transformed state of the original $|\phi_{\mathbf{k}} \rangle \equiv |\psi_{\mathbf{k}} \rangle \otimes |S_{\mathbf{k}} \rangle \otimes |P_{\mathbf{k}} \rangle$ is given by
\bea\label{state}
|\widetilde{\phi}_{\mathbf{k}}(\xi, \eta,  t)\rangle_{WZ} = |\widetilde{\psi}_{\mathbf{k}}(\xi,  t)\rangle \otimes |\widetilde{S}_{\mathbf{k}}(\eta,  t)\rangle \otimes |\widetilde{P}_{\mathbf{k}}(\eta,  t)\rangle\,,
\eea
 where the transformed fermion and boson states are
$
|\widetilde{\psi}_{\mathbf{k}}(\xi,  t)\rangle =\widetilde{\alpha}^{r\dag}_{\mathbf{k}}(\xi,  t)\rangle|\widetilde{0}(\xi,  t)\rangle_{\psi} = J^{-1}_{\psi}(\xi,  t)|\psi_{\mathbf{k}} \rangle  $,
$ |\widetilde{S}_{\mathbf{k}}(\eta,  t)\rangle  = \widetilde{b}^{\dag}_{\mathbf{k}}(\xi,  t)\rangle|\widetilde{0}(\xi,  t)\rangle_{S}= J^{-1}_{S}(\eta,  t)|S_{\mathbf{k}}\rangle $ and
$  |\widetilde{P}_{\mathbf{k}}(\eta,   t)\rangle = \widetilde{c}^{\dag}_{\mathbf{k}}(\eta,  t)\rangle|\widetilde{0}(\eta,  t)\rangle_{P}= J^{-1}_{P}(\eta,  t)|P_{\mathbf{k}}\rangle
$. For notational simplicity we will use $|\widetilde{\phi}_{\mathbf{k}}(  t)\rangle_{WZ} \equiv |\widetilde{\phi}_{\mathbf{k}}(\xi, \eta,  t)\rangle_{WZ} = J^{-1}(\xi,\eta,  t)|\phi_{\mathbf{k}} \rangle\,$.

The generators $J$s are unitary operators if $\mathbf{k}$ assumes a discrete  range of values; the Fock spaces built on the  states $|0\rangle $ and $|\tilde{0}(t)\rangle$ are then unitarily equivalent. On the other hand, in the  $\mathbf{k}$  continuous limit, i.e. in QFT, the $J$s are not unitary operators \cite{Takahashi:1974zn}. This implies that $|\tilde{0}(t) \rangle$ cannot be expressed as a superposition of states belonging to the Fock space built over $|0\rangle$. Rather, this last one and the Fock space generated by repeated applications of $\tilde{\alpha}^{r\,\dag}_{\mathbf{k}}( \xi,  t)$, $\tilde{b^{\dag}}_{\mathbf{k}}(\eta,  t)$ and $\tilde{c}^{\dag}_{\mathbf{k}}(\eta,  t)$  over $|\tilde{0}(t)\rangle$ are unitarily inequivalent, and therefore physically different state spaces.\\
\indent When one considers systems characterized by the presence of condensates,  such as those mentioned above  \cite{Unruh:1976db,Schwinger:1951nm,Bardeen:1957mv,Takahashi:1974zn,Celeghini:1991yv},
the state $|\tilde{0}( t)\rangle$  is then the relevant physical vacuum with  the nontrivial structure of a condensate~\cite{Capolupo:2012vf}
responsible of its non-zero energy density and of the appearance of the AAIs~\cite{Capolupo:2013fasi}.

Let  $H =  H_B + H_{\psi} $ denote the free Hamiltonian corresponding to the Lagrangian (\ref{WS}),  with $H_B = H_S + H_P$. One finds $ \langle\tilde{0}(t)| H_B |\tilde{0}(  t)\rangle = \sum_{\mathbf{k}}\, \omega_{\mathbf{k}} (1 + 2 |V^{B}_{\textbf{k}}(\eta,t)|^2)$
for bosons and $
\langle\tilde{0}(  t)| H_{\psi} |\tilde{0}(  t)\rangle = - \sum_{\mathbf{k}}\, \omega_{\mathbf{k}} \,(1- 2 |V^{\psi}_{\mathbf{k}}(\xi,t)|^2)$ for fermions
 and the expectation value of  $H$ on  $|\tilde{0}( t)\rangle$ is non-vanishing and positive
and represents the background noise \cite{Capolupo:2012vf}
\bea\label{Ht}
\langle\tilde{0}(  t)| H |\tilde{0}(  t)\rangle = 2 \sum_{\mathbf{k}}\, \omega_{\mathbf{k}} (|V^{B}_{\textbf{k}}(\eta,t) |^2 + |V^{\psi}_{\textbf{k}}(\xi,t)|^2 ) ,
\eea
which implies the spontaneous breakdown of SUSY \cite{Capolupo:2012vf}.
The physical system under study determines the symmetry breakdown conditions through the Bogoliubov coefficients $V^{B}_{\textbf{k}}$ and  $V^{\psi}_{\textbf{k}}$. Their physical meaning in turn depends on the specific meaning attached to the Bogoliubov transformation parameters.
For example, in the case of thermal field theories (e.g. in the Thermo Field Dynamics formalism), the transformation parameters  are related to  temperature, the physical vacuum is the thermal one, and the result is that SUSY is spontaneously broken at any nonzero temperature, as it is well known \cite{Das:1997gg,Buchholz:1997mf}.
\\
\indent {\it Aharonov-Anandan invariant} -- The systems in which the vacuum condensate is physically relevant also have the common characteristic of an AAI \cite{Anandan:1990fq} in their time  evolution~\cite{Capolupo:2013fasi}. Such an invariant could be used as a tool to reveal the SUSY breakdown. In general, to generate an AAI in the evolution of a state $|\chi_{\mathbf{k}}(t)\rangle$  it is necessary and sufficient that its  energy uncertainty $\Delta E_{\mathbf{k}} ^{2}(t) = \langle \chi_{\mathbf{k}}(t)|H^{2}|\chi_{\mathbf{k}}(t)\rangle -  (\langle \chi_{\mathbf{k}}(t)|H|\chi_{\mathbf{k}}(t)\rangle)^{2}$   be non-vanishing, which happens, indeed, in the cases above considered~\cite{Unruh:1976db}--\cite{Celeghini:1991yv}.
The Aharonov--Anandan invariant  is then defined as
$S_{\mathbf{k}} = \frac{2}{\hbar} \int_{0}^{  t}  \Delta E_{\mathbf{k}} (t^{\prime}) \, dt^{\prime}\,$.

For each of the neutral scalar and fermion fields we refer to, e.g. in Eq.~(\ref{state}), the energy variances due to the Bogoliubov  coefficients are $\Delta E_{\mathbf{k}}^{S}(t) = \Delta E_{\mathbf{k}} ^{P}(t)=\sqrt{2} \omega_{\bf k} |U^B_{\bf k}(\eta,t)| |V^B_{\bf k}(\eta,t)|,$ and
$\Delta E_{\mathbf{k}}^{\psi}(t) =   \omega_{\bf k} |U^\psi_{\bf k}(\eta,t)| |V^\psi_{\bf k}(\eta,t)|,$ respectively.
The corresponding AAIs are then
\bea\label{AAI-Bog-Bos}
 S_{\mathbf{k}}^{S}(t) = S_{\mathbf{k}}^{P}(t)\,=2\,\sqrt{2}\int_{0}^{  t} \omega_{\bf k} |U^B_{\bf k}(\eta,t^{\prime})|| V^B_{\bf k} (\eta,t^{\prime})|\,dt^{\prime}\,,
\eea
for scalar and pseudoscalar bosons  and
\bea\label{AAI-Bog-Fer}
S_{\mathbf{k}}^{\psi}(t)\,=\,2 \int_{0}^{  t}\omega_{\bf k} |U^\psi_{\bf k}(\xi,t^{\prime})| |V^\psi_{\bf k}(\xi,t^{\prime})|\, dt^{\prime},
\eea
for Majorana fermion field.

In the case of the free Wess-Zumino model, the AAI associated to the Bogoliubov transformed state $|\widetilde{\phi}_{\mathbf{k}}(\xi,\eta,  t)\rangle_{WZ} $ (Eq.(\ref{state}))
 is then
\begin{widetext}
\bea\label{DeltaPhase}
S_{\mathbf{k}}^{WZ} (t) &=& 2 \int_{0}^{  t}\omega_{\bf k}\sqrt{4|U^B_{\bf k}(\eta,t^{\prime})|^{2} |V^B_{\bf k} (\eta,t^{\prime})|^{2}
\,+\,
 |U^\psi_{\bf k}(\xi,t^{\prime})|^{2} |V^\psi_{\bf k}(\xi,t^{\prime})|^{2}} dt^{\prime}.
\eea
\end{widetext}

The non-vanishing value of $S_{\mathbf{k}}^{WZ} (t)$ is due to the presence of the condensates and controlled by the Bogoliubov coefficients responsible of the SUSY breakdown. We thus see that in the presence of a condensate, the invariant $S_{\mathbf{k}}^{WZ} (t)$ signals SUSY breakdown. $S_{\mathbf{k}}^{WZ} (t)$ is indeed zero when the condensates disappear (SUSY restoration).\\
  \indent These results  apply to any phenomena in which condensates are present.
To our knowledge, the only system that deserves a separate discussion is the particle mixing phenomenon.
Although spontaneous breakdown of SUSY is present in the field mixing~\cite{Capolupo:2010ek,Mavromatos:2010ni}, it would be hardly revealed through an AAI measurement. It is so because here
the AAI arises as an effect of the mixing of fields with the addition of a small contribution due to the condensate structure~\cite{Blasone:2009xk,
Capolupo:2011rd}. Since this last one is not affecting much the value of the AAI, the SUSY breaking is difficult to be revealed. In the other  systems above considered, on the contrary, the AAI is solely due to the condensate contribution. This fact makes the difference between these last cases and the field mixing.

We now analyze the specific case of  thermal states and propose a possible experiment to detect SUSY violation with AAI.\\
\indent {\it Thermal state} -- In the  Thermo Field Dynamics formalism  \cite{Takahashi:1974zn} one can show that the thermal vacuum is a condensate generated through Bogoliubov transformations whose  parameter is related to temperature. We refer for details to the existing literature, see e.g.  \cite{Takahashi:1974zn}. Here we only recall that the Bogoliubov coefficients $U$ and $V$  are real and given by
$U_{{\bf k}}(\theta) = \sqrt{\frac{e^{\beta\hbar \omega_{\bf k} }}{e^{\beta\hbar \omega_{\bf k } }\pm 1}}$ and
$V_{{\bf k}}(\theta) = \sqrt{\frac{1}{e^{\beta\hbar \omega_{\bf k} } \pm 1}}$, with $-$ for bosons and $+$ for fermions, and
 $\beta = 1/ k_{B}T$.

 The uncertainties in the energy of the temperature dependent state
$
|\widetilde{\phi}_{\mathbf{k}}(\theta)\rangle\, \equiv \alpha_{\mathbf{k}}^{\dag}(\theta)|0(\theta)\rangle\,,
$ with $\theta \equiv (\theta , t)$,
are given by
\bea\non
\Delta E_{\mathbf{k}}^{S}(\theta) = \Delta E_{\mathbf{k}}^{P}(\theta) &=&\sqrt{2}\,\hbar \omega_{\bf k} \, U^{B}_{\bf k}(\theta) V^{B}_{\bf k}(\theta)
\\
&=& \sqrt{2}\hbar \omega_{\bf k} \,\frac{{e^{\beta \hbar \omega_{\bf k}/2  }}}{{(e^{\beta \hbar \omega_{\bf k } }-1)}}\,,
\eea
for bosons, and
\bea\non
\Delta E_{\mathbf{k}}^{\psi}(\theta) = \hbar \omega_{\bf k} \, U^{\psi}_{\bf k}(\theta) V^{\psi}_{\bf k}(\theta)
=  \hbar \omega_{\bf k} \,\frac{{e^{\beta \hbar \omega_{\bf k}/2  }}}{{(e^{\beta \hbar \omega_{\bf k } }+1)}}\,,
\eea
for the Majorana fermion field.
 The corresponding AAIs are
\bea
S_{\mathbf{k}}^{S}(\theta,t) &=& S_{\mathbf{k}}^{P}(\theta,t) = 2\sqrt{2}\, \omega_{\bf k}\, t\, \frac{e^{ \beta\hbar \omega_{\bf k}/2  }}{e^{\beta\hbar \omega_{\bf k } } - 1}\,,
\\
S_{\mathbf{k}}^{\psi}(\theta,t) &=& 2 \,\omega_{\bf k}\, t\, \frac{e^{ \beta\hbar \omega_{\bf k}/2  }}{e^{\beta\hbar \omega_{\bf k } } + 1}\,.
\eea
In the supersymmetric model, at $T\neq 0$, SUSY is always broken and the masses are thermally renormalized \cite{Midor,Midor2,Midor3} assuming different values for the fields $\psi$, $S$ and $P$, i.e. $m_\psi \neq m_P \neq m_S $. The dispersion relations for the three particles are different and then three different AAIs arise for $\psi$, $S$ and $P$ fields, $S_{\mathbf{k}}^{\psi} \neq S_{\mathbf{k}}^{S} \neq S_{\mathbf{k}}^{P}$.
On the contrary, at $T=0$, the AAIs are vanishing, $S_{\mathbf{k},T = 0}^{\psi} = S_{\mathbf{k},T = 0}^{S}= S_{\mathbf{k},T = 0}^{P} = 0$, since the Bogoliubov coefficients are zero and, as well known, SUSY is preserved.

The presence of the AAI for finite temperature systems  signals SUSY breakdown.
An experiment has been proposed recently in which the Wess-Zumino model in $2+1$ dimensions can emerge from a mixture of cold atoms and molecules trapped in two dimensional optical lattices \cite{Yue-Yang}.
In this system the superpartner of the fermion atom is represented by a bosonic diatomic molecule and SUSY is preserved at zero temperature. At non-zero temperature SUSY is broken and a signature of such a violations can be probed  by detecting a thermal Goldstone fermion, the phonino.
The detection of such a particle is complicated. However, inspired by the method of \cite{Yue-Yang}, in order to partially bypass such difficulties, we propose to construct an experimental setup as follows. Our setting is aimed to detect the AAI's and behaves like an interferometer where the AAI results from the different temperature properties  of two similar atom-molecule systems.  In other words, the apparatus provides the measure of
the difference between the geometric invariants of two mixtures of cold atoms and molecules trapped in two coplanar, two-dimensional optical lattices, one at temperature $T\neq 0$ and the other one at $T = 0$. These lattices are moved along the direction perpendicular to the common lattice plane. The part of the apparatus relative to the lattice at $T\neq 0$  has non-vanishing invariant
\bea\non\label{WZ-AAI}
S_{\mathbf{k}}^{WZ}(\theta,t) = \sqrt{2}  \omega_{\bf k}  t \,\sqrt{[3+5\cosh (\beta\hbar \omega_{\bf k})] \textrm{csch}^{2}(\beta\hbar \omega_{\bf k})}\,.
\\
\eea
 The other one, at $T=0$, has $S_{\mathbf{k},T=0}^{WZ}=0$. Use of such an experimental setup could lead to plots similar to the ones theoretically obtained in Figs.~1,~2, and 3. These are derived by computing the AAIs {\it vs} the excitation energy for different temperatures of the condensate. In particular, we have considered atomic excitation frequencies characteristic of Bose-Einstein condensate of order of $ 2 \times 10^{4} s^{-1}-10^{5} s^{-1} $  and temperatures of the condensate of the order of $(20-200) nK$. For simplicity, we considered the case of equal masses of fermion and boson fields, $m_\psi = m_P = m_S$, which happens in the massless version of Wess-Zumino model also presented in \cite{Yue-Yang}. In this case, the AAIs for scalar and pseudoscalar fields are the same, $S_{\mathbf{k}}^{P} = S_{\mathbf{k}}^{S} $. The generalization to the massive Wess-Zumino model is straightforward.
We find that, in principle, the invariants reported in the above pictures are all detectable.
\begin{figure}
\begin{picture}(300,180)(0,0)
\put(10,20){\resizebox{9 cm}{!}{\includegraphics{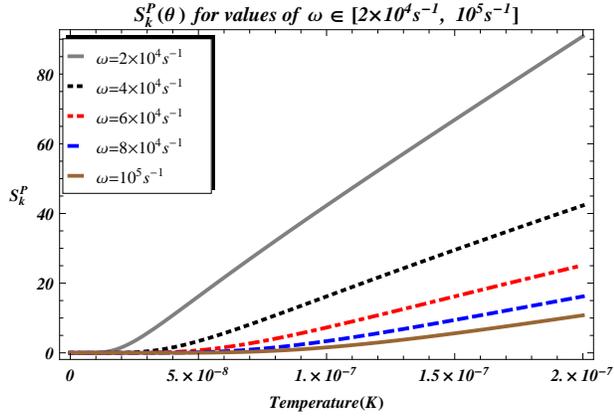}}}
\end{picture}\vspace{-1cm}
\caption{\em Plots of AAI for bosons, $S_{\mathbf{k}}^{P}$
as a function of temperature $T$, for a time interval $t = 4 \times 2\pi /\omega $ and for sample values of  $\omega \in  [2 \times 10^{4} s^{-1},  10^{5} s^{-1}]$, as indicated in the inset. We consider the massless case and we assume that $ S_{\mathbf{k}}^{P} = S_{\mathbf{k}}^{S}$. We neglect time dependence of $T$.}
\label{pdf}
\end{figure}
\begin{figure}
\begin{picture}(300,180)(0,0)
\put(10,20){\resizebox{9 cm}{!}{\includegraphics{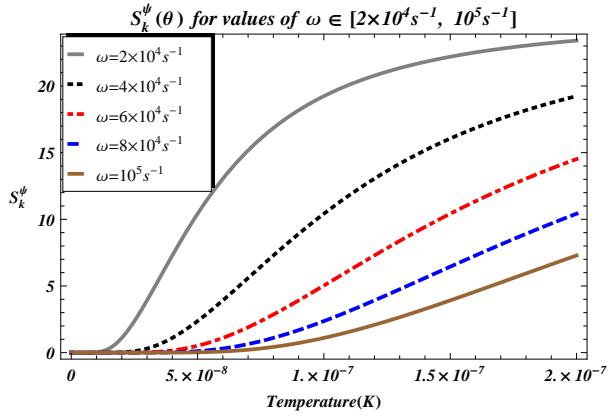}}}
\end{picture}\vspace{-1cm}
\caption{\em Plots of AAI for fermions, $S_{\mathbf{k}}^{\psi}$
as a function of temperature $T$, for the same time interval and sample values of  $\omega$   as  in Fig.1.}
\label{pdf}
\end{figure}
\begin{figure}
\begin{picture}(300,180)(0,0)
\put(10,20){\resizebox{9 cm}{!}{\includegraphics{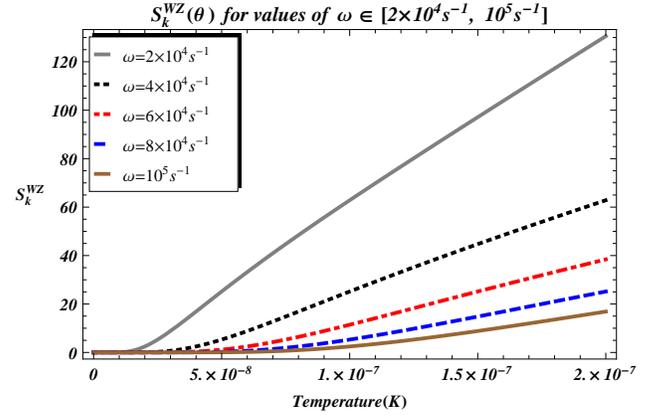}}}
\end{picture}\vspace{-1cm}
\caption{\em Plots of the total AAI induced by SUSY breaking, $ S_{\mathbf{k}}^{WZ}$,
as a function of temperature $T$, for the same time interval and sample values of  $\omega$  as  in Figs.1 and 2.}
\label{pdf}
\end{figure}

From Fig.~2 we see that the fermionic AAI shows a plateau at higher values of the temperature parameter. On the other hand, at high temperature (above $\thickapprox 200 ~nK$) also the condensate is expected to disappear (and with it the AAI vanishes).

Finally, we note that, also
in the supersymmetric context with a condensate,
the noise does not affect AAIs, as happens
for any system which presents a condensate structure~\cite{Capolupo:2013fasi}.
Let us demonstrate that for the supersymmetric state given in Eq.(\ref{state}).

In the case of free Wess-Zumino model, the background  noise is due to the nonzero energy of the physical vacuum as shown in Eq.(\ref{Ht}). This contribution, responsible of the SUSY breaking is, in general, a c-number.
For example, in the particular case of the two-dimensional optical lattice of cold atoms and molecules, simulating the massless Wess-Zumino model, the thermal noise is a constant given by
$
\langle\tilde{0}(t)| H |\tilde{0}(t)\rangle =\,14 \, \pi\,\zeta(3)\, T^{3}\,.
 $

As well known, a c-number does not modify the uncertainty $\Delta E_{\mathbf{k}}(t) $ in the energy of the system, thus the background noise does not modify the AAI.
 Indeed, let us denote the noise with $\Lambda = \langle\tilde{0}(t)| H |\tilde{0}(t)\rangle $ and take into account also $\Lambda$ in the computation of  the uncertainty in the energy. For the Bogoliubov transformed state $|\widetilde{\phi}_{\mathbf{k}}(t)\rangle_{WZ}\,$ (\ref{state}) it is immediate to verify that $\Delta E_{\mathbf{k}} (t)$ is invariant under the translation $H \rar H + \Lambda$:
  \bea\non
 \Delta E^{2}_{\mathbf{k}} (t) &=& _{WZ}\langle \widetilde{\phi}_{\mathbf{k}}(t)|(H \,+\, \Lambda)^{2}|\widetilde{\phi}_{\mathbf{k}}(t)\rangle_{WZ}
 \\ \non
  &-&  (_{WZ}\langle \widetilde{\phi}_{\mathbf{k}}(t)|(H \,+\, \Lambda)|\widetilde{\phi}_{\mathbf{k}}(t)\rangle_{WZ})^{2}
\\[5pt]\non
 &=& _{WZ}\langle \widetilde{\phi}_{\mathbf{k}}(t)| H^{2}|\widetilde{\phi}_{\mathbf{k}}(t)\rangle_{WZ}
 \\
  &-&  (_{WZ}\langle \widetilde{\phi}_{\mathbf{k}}(t)| H |\widetilde{\phi}_{\mathbf{k}}(t)\rangle_{WZ})^{2}.
  \eea
This fact leaves unchanged the AAI (up to temperatures above which the AAI does not disappear by itself, as commented above). Note that, on the contrary, the noise affects Berry-like phases since $\Lambda$ contributes to the value of the Hamiltonian for the system evolution.

In conclusion, we have shown that, in a supersymmetric field theory model, the SUSY spontaneous  breakdown induced in different phenomena by the  vacuum condensates can be revealed by  the AAI.
The breakdown of SUSY due to the non-zero energy of the vacuum condensate is indeed related to the Bogoliubov coefficients which also appear  in the AAI.  The analysis is presented in the case of the Wess-Zumino model and might be exploited in
a recently proposed experimental setup based on optical lattices in which SUSY is broken at
non-zero temperature.
Indeed, the difference of the geometric invariants between the AAIs of a system in a thermal bath and one at zero temperature
signals SUSY breaking in such a system.

\section*{Acknowledgements}
Partial financial support from MIUR and INFN is acknowledged.

\section*{Conflict of Interests}
The authors declares that there is no conflict of interests regarding the publication of this article.

\end{document}